\documentclass[sigconf]{acmart}
\AtBeginDocument{%
  \providecommand\BibTeX{{%
    \normalfont B\kern-0.5em{\scshape i\kern-0.25em b}\kern-0.8em\TeX}}}

\setcopyright{acmlicensed}
\copyrightyear{2018}
\acmYear{2018}
\acmDOI{XXXXXXX.XXXXXXX}

\acmConference[Conference acronym 'XX]{Make sure to enter the correct
  conference title from your rights confirmation emai}{June 03--05,
  2018}{Woodstock, NY}
\acmISBN{978-1-4503-XXXX-X/18/06}

\usepackage{multirow}
\usepackage{enumitem}
\usepackage{graphicx}
\usepackage{caption}
\usepackage{subfigure}
\usepackage{amsthm}

\copyrightyear{2024}
\acmYear{2024}
\setcopyright{rightsretained}
\acmConference[CIKM '24]{Proceedings of the 33rd ACM International Conference on Information and Knowledge Management}{October 21--25, 2024}{Boise, ID, USA}
\acmBooktitle{Proceedings of the 33rd ACM International Conference on Information and Knowledge Management (CIKM '24), October 21--25, 2024, Boise, ID, USA}
\acmDOI{10.1145/3627673.3679593}
\acmISBN{979-8-4007-0436-9/24/10}

\begin{document}

\title{Combining Incomplete Observational and Randomized Data \\ for Heterogeneous Treatment Effects}

\author{Dong Yao}
\affiliation{%
  \institution{Ant Group}
  \city{Hangzhou}
  \country{China}}
\email{yaodong.yao@antgroup.com}

\author{Caizhi Tang}
\affiliation{%
  \institution{Ant Group}
  \city{Hangzhou}
  \country{China}
}
\email{caizhi.tcz@antgroup.com}

\author{Qing Cui}
\affiliation{%
  \institution{Ant Group}
  \city{Hangzhou}
  \country{China}
}
\email{cuiqing.cq@antgroup.com}

\author{Longfei Li}
\authornote{Corresponding Authors.}
\affiliation{%
  \institution{Ant Group}
  \city{Hangzhou}
  \country{China}
}
\email{longyao.llf@antgroup.com}

\renewcommand{\shortauthors}{Dong Yao, Caizhi Tang, Qing Cui, and Longfei Li}
\newcommand{\etal}{\textit{et al}.}
\newcommand{\ie}{\textit{i}.\textit{e}.}
\newcommand{\eg}{\textit{e}.\textit{g}.}
\newcommand{\vpara}[1]{\vspace{0.05in}\noindent\textbf{#1 }}

\theoremstyle{plain}
\newtheorem{mytheorem}{Theorem}[section]
\newtheorem{assumption}[theorem]{Assumption}
\theoremstyle{remark}
\newtheorem{remark}{Remark}
\begin{abstract}

Data from observational studies (OSs) is widely available and readily obtainable yet frequently contains confounding biases. On the other hand, data derived from randomized controlled trials (RCTs) helps to reduce these biases; however, it is expensive to gather, resulting in a tiny size of randomized data. For this reason, effectively fusing observational data and randomized data to better estimate heterogeneous treatment effects (HTEs) has gained increasing attention. However, existing methods for integrating observational data with randomized data must require \textit{complete} observational data, meaning that both treated subjects and untreated subjects must be included in OSs. This prerequisite confines the applicability of such methods to very specific situations, given that including all subjects, whether treated or untreated, in observational studies is not consistently achievable. In our paper, we propose a resilient approach to \textbf{C}ombine \textbf{I}ncomplete \textbf{O}bservational data and randomized data for HTE estimation, which we abbreviate as \textbf{CIO}. The CIO is capable of estimating HTEs efficiently regardless of the completeness of the observational data, be it full or partial. Concretely, a confounding bias function is first derived using the pseudo-experimental group from OSs, in conjunction with the pseudo-control group from RCTs, via an effect estimation procedure. This function is subsequently utilized as a corrective residual to rectify the observed outcomes of observational data during the HTE estimation by combining the available observational data and the all randomized data. To validate our approach, we have conducted experiments on a synthetic dataset and two semi-synthetic datasets. 
  
\end{abstract}

\ccsdesc[300]{Computing methodologies~Machine learning}
\ccsdesc[500]{Computing methodologies~Causal reasoning and diagnostics}
\ccsdesc[500]{Mathematics of computing~Causal networks}

\ccsdesc[500]{Computing methodologies~Machine learning}

\keywords{Causal Inference, Heterogeneous Treatment Effects, Observational Data, Random Control Trial Data}




\maketitle

\section{Introduction}
\begin{figure}[t]
    \centering
    \includegraphics[width=\linewidth]{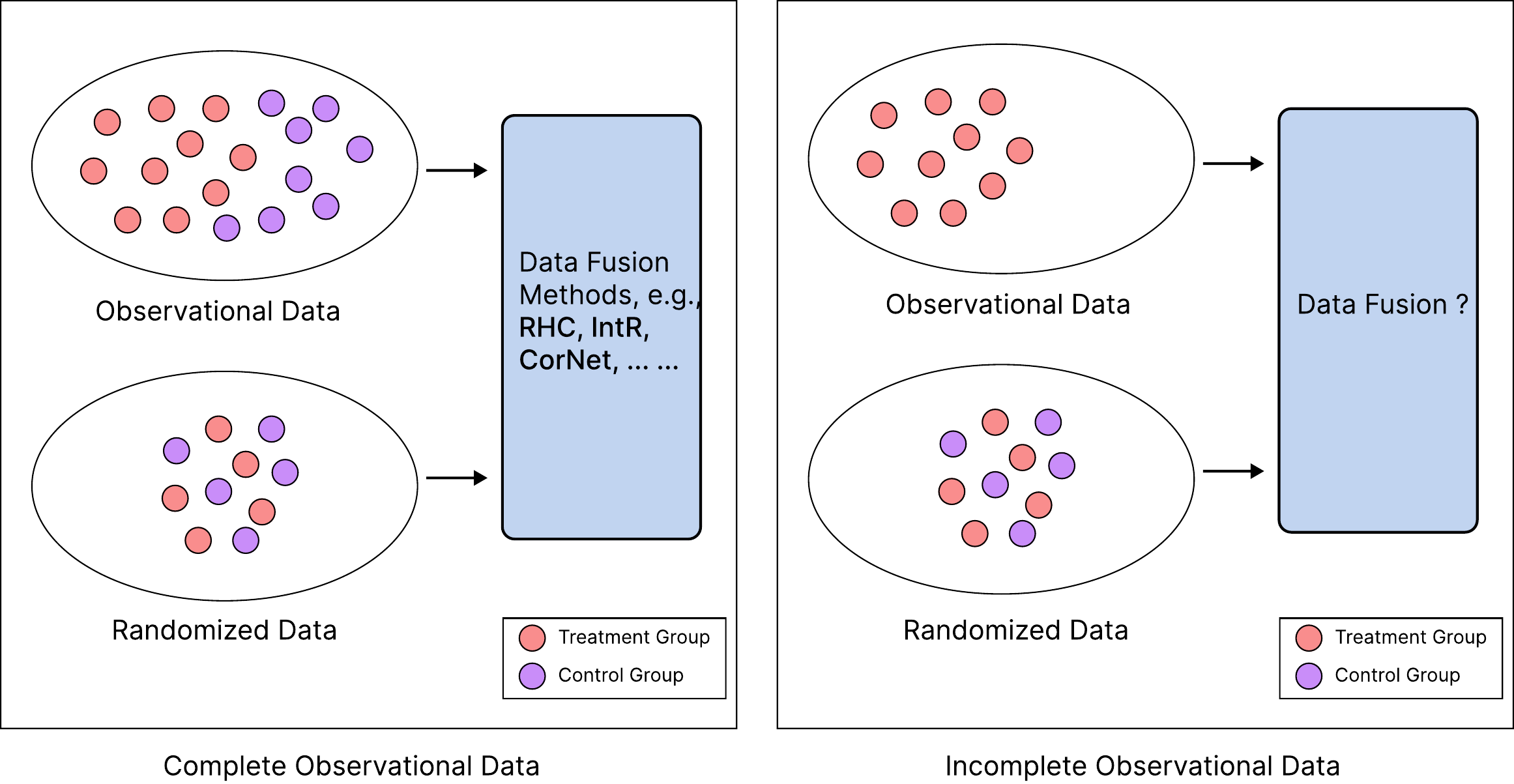}
    \caption{The data composition under the two situation: complete and incomplete OS data. For illustration, the right subfigure demonstrates a case where the control group is missing. It should be noted that in practice, the treatment group could also be absent. }
    \label{fig:instance}
\end{figure}
Heterogeneous treatment effects (HTEs) refer to the variations in causal effects of a treatment or an intervention across different sub-populations, based on their distinct characteristics or contexts. It is of great importance to estimate HTEs in various fields, such as medicine \cite{hamburgPathPersonalizedMedicine2010,glassCausalInferencePublic2013}
, marketing \cite{brodersenInferringCausalImpact2015,atheyPredictionUsingBig2017} and epidemiology \cite{robinsMarginalStructuralModels2000}.

There are two types of data in causal inference: \texttt{observational data} and \texttt{randomized data}. 
Observational data is collected without any intervention from observational studies (OSs), reflecting real-world conditions as they naturally occur.
Considering the advantages of low-cost of acquisition and vast quantities, most existing literature \cite{atheyGeneralizedRandomForests2018,johanssonLearningRepresentationsCounterfactual,kunzelMetalearnersEstimatingHeterogeneous2019,nieQuasiOracleEstimationHeterogeneous2020,powersMethodsHeterogeneousTreatment2018,shalitEstimatingIndividualTreatment2017,wagerEstimationInferenceHeterogeneous2017} focus on estimating HTEs from observational data.
Although it is valuable for its real-world relevance and the volume of data it can provide, observational data is often prone to confounding biases that can challenge causal interpretations. In practice, those methods mentioned above depend on some assumptions,e.g., assuming the absence of unobserved confounders, which is not testable and difficult to satisfy in practice.
For instance, when doctors prescribe medication, they consider various patient-specific factors, some of which may not be captured in the medical records. Relying solely on observational data in this case can result in confounding bias, since unrecorded influences on both the treatment decisions and outcomes remain unaccounted for. This leads to challenges in identifying HTEs accurately and can introduce bias into the estimates of treatment effects. Therefore, it is unreasonable to make use of observational data for HTE estimation in practice without making the unconfoundedness assumption.

Randomized data, particularly from Randomized Controlled Trials (RCTs), is generated in a randomized controlled experimental setting where participants are randomly assigned to different groups to isolate the effect of a treatment or intervention. And the trial data is deemed the gold standard for estimating HTEs. However, trial data is often limited by the costs, laws, and ethics. Taking medicine as an example, it's impossible to conduct large-scale clinical trials, especially for those side-effect drugs. 

Considering the unique attributes of RCTs and OSs, the integration of data from both sources has gained traction as a method to estimate HTEs. References such as \cite{atheyCombiningExperimentalObservational2020, chengAdaptiveCombinationRandomized2021, ghassamiCombiningExperimentalObservational2022, guFASTFusedAccurate2023, hattCombiningObservationalRandomized2022, kallusRemovingHiddenConfounding2018, rosenmanCombiningObservationalExperimental2020, yangCombiningMultipleObservational2021, yangIntegrativeLearnerHeterogeneous2022} highlight this growing trend. Nevertheless, the prevalent techniques for HTE estimation largely hinge on the availability of \textit{complete} observational datasets. Certain methods begin by constructing an HTE estimator exclusively based on observational data \cite{kallusRemovingHiddenConfounding2018, chengAdaptiveCombinationRandomized2021, hattCombiningObservationalRandomized2022}, whereas others necessitate the creation of a propensity score model derived from such data \cite{yangIntegrativeLearnerHeterogeneous2022}. For example, RHC \cite{kallusRemovingHiddenConfounding2018} supposes that the confounding bias is a parametric function that can be learned. It learns a biased estimator by training on observational data, and then uses randomized data to remove the bias. In another instance, \citet*{yangIntegrativeLearnerHeterogeneous2022} also adopts a parametric approach to model the confounding bias when estimating HTEs. They achieve this by using an integrative $R$-learner (IntR) that merges data from RCTs and OSs. Additionally, this method necessitates a pre-learned propensity score estimator based on both randomized and observational data. Nonetheless, OSs often suffer from incompleteness due to the intricate nature of real-world scenarios. \textit{The complete observational data is a dataset which contains control data and treatment. Therefore, when the one of control group and treatment group from observational data is absent, we refer it as incomplete observational data}. We draw the Figure \ref{fig:instance} to demonstrate the difference between two situation. As an illustration, consider a scenario where an new experimental drug is introduced to treat a chronic illness such as diabetes. Due to the drug's recent entry into the market and lack of extensive track record, patients might be skeptical about its benefits and potential side effects. Consequently, many of them may choose to stick with their current treatment regimens rather than try the new medication. This reluctance can result in scarce or incomplete data for the treatment group within the OS, as the majority of patients remain within the untreated or control cohorts. In other words, it is possible to gather data from patients who have not been treated with the new experimental drug (control data in OSs), while lacking data from patients who have undergone treatment (treatment data in OSs). Under such circumstances, traditional data fusion approaches are ill-suited for assessing the impact of the new experimental drug. The deficiency of treatment group or control group in OSs can result in a considerable decrease in the efficacy of these methods or may even lead to their failure to function.

In this paper, we introduce a robust technique, termed \textbf{CIO}, designed to integrate incomplete observational data with randomized trial data for estimating HTEs. Besides, the versatility of CIO extends to the combination of \textit{complete} observational data with randomized data for HTE estimation. Our approach overcomes the limitations of current data fusion methodologies, which require complete observational data for HTE estimation, thus offering a more practical solution for real-world scenarios. 
Initially, by creating a dummy treatment, we designate the treatment group or control group of observational as a pseudo-experimental group and the all randomized data is used to constitute the pseudo-control group. Subsequently, we utilize the learning pattern of effect estimation to learn a confounding bias function using the pseudo-experimental group and the pseudo-control group. Finally, we integrate the entirety of the data at our disposal, including the available observational data and the all randomized data, to derive an HTE estimator. Simultaneously, we perform debiasing for observational data, with the assistance of the confounding bias function, which is employed as a residual to correct for the observed outcomes of observational data. The specific training process and details are illustrated in Section \ref{method}. The main contributions of our work are as follows:
\begin{itemize}[leftmargin=*]
    \item We introduce a robust approach, termed CIO, that fully leverages the strengths of both observational data and randomized data while addressing the shortcomings of existing data fusion techniques, i.e., their reliance on complete observational data for estimating HTEs.
    \item We form pseudo-experimental and pseudo-control group from another perspective to train an estimator, which is intended to serve as a confounding bias function. This is an innovative tactic in the assessment of confounding bias.
    \item We validate our approach through extensive experimentation on one synthetic dataset and two real-world datasets, demonstrating that our method not only combines observational data and randomized data more effectively for HTE estimation but also retains its efficacy in scenarios where the data from OSs is partially missing.
\end{itemize}
\section{Related Work}

\subsection{Heterogeneous Treatment Effect Estimation}
Accurately estimating heterogeneous treatment effects is considerably significant for medicine, marketing, epidemiology and other related areas. For that reason, vast machine learning methods have been proposed to estimate HTEs. We classify these methods into three category: tree-based methods \cite{atheyGeneralizedRandomForests2018,wagerEstimationInferenceHeterogeneous2017}, bayesian algorithms \cite{alaaBayesianInferenceIndividualized2017,alaaLimitsEstimatingHeterogeneous2018,zhangLearningOverlappingRepresentations2020} and deep learning algorithms \cite{johanssonLearningRepresentationsCounterfactual,shalitEstimatingIndividualTreatment2017,yaoRepresentationLearningTreatment2018,yoonGANITEESTIMATIONINDIVIDUALIZED2018}. However, these methods all make a strong unconfounding premise for observational data, which can not be verified and often does not stand up to real-world scrutiny. As a result, this prevents the aforementioned techniques from being implemented in practical settings. To solve this common problem, certain techniques \cite{kuzmanovicDeconfoundingTemporalAutoencoder2021,louizosCausalEffectInference2017a} aim to identify the actual confounders by working with noisy indicators that serve as proxies for these confounders. Yet, it remains uncertain if the covariates we observe genuinely act as surrogates for the actual confounding variables. Other strategies aim to infer missing confounders through the assignment data from multiple treatments (Wang and Blei, 2019; Bica et al., 2019) or over time with treatments administered in sequence (Hatt and Feuerriegel, 2021b). Nevertheless, these methods also rely on hypotheses such as single strong ignorability and constancy of confounders over time, which presume an absence of hidden confounders. Since these suppositions are not verifiable in real-world applications, the practical applicability of such techniques is impeded. GBCT \cite{tangDebiasedCausalTree2022} is another attemp to integrate current observational data and their historical controls.

\subsection{Combining Observational and Randomized Data}
Recently, merely a few methods \cite{atheyCombiningExperimentalObservational2020,chengAdaptiveCombinationRandomized2021,ghassamiCombiningExperimentalObservational2022,guFASTFusedAccurate2023,hattCombiningObservationalRandomized2022,kallusRemovingHiddenConfounding2018,rosenmanCombiningObservationalExperimental2020,yangCombiningMultipleObservational2021,yangIntegrativeLearnerHeterogeneous2022} have been proposed to combine observational data and randomized data for estimating HTEs. The RHC approach, as described by \cite{kallusRemovingHiddenConfounding2018}, assumes that confounding bias can be represented as a learnable parametric function. It involves training a biased estimator with observational data and subsequently using data from randomized trials to correct the confounding bias. In a related vein, \cite{chengAdaptiveCombinationRandomized2021} suggest obtaining one estimate from observational data and another from randomized data, subsequently combining these two estimates through a weighted average. Nonetheless, the process of calibrating the weights for this averaging necessitates a substantial randomized data validation set. Our experimental observations indicate that this requirement is at odds with the typically limited size of randomized data samples. \cite{yangIntegrativeLearnerHeterogeneous2022} parametrically formulates the confounding bias function and an effect estimator for HTE analysis, leveraging an integrative $R$-learner that fuse data from RCTs and OS. \cite{hattCombiningObservationalRandomized2022} introduces CorNet—a dual-phase framework that exploits a common structural aspect of both data kinds. More recently, FAST \cite{guFASTFusedAccurate2023}, as a tree-based method, draws from the statistical principle of shrinkage estimation. It crafts a weighting strategy that is optimized to strike a balance between the unbiased estimator derived from trial data and the estimator from observational data, which may carry bias. 
\section{Method} \label{method}
In this section, our goal is to provide a comprehensive introduction to the proposed CIO method. We begin by presenting the foundational elements, which encompass the definition of variables and the essential assumptions required for our approach. Following that, we will illustrate the training steps involved in CIO and discuss the specific training loss function employed in the process. Finally, we will give a in-depth analysis in theoretical about why our proposed method is effective.

\subsection{Preliminary} \label{method:pre}
In our study, we concentrate on a scenario where the treatment variable $T$ is binary, taking values in the set $\{0, 1\}$. We denote $\mathbf{X} \in \mathbb{R}^p$, as the vector of covariates measured before being treated, and $Y \in \mathbb{R} $ as the outcome variable of interest. Employing the potential outcomes framework \cite{rubinEstimatingCausalEffects1974}, we define causal effects using $Y(t)$ to represent the potential outcome if the subject were to receive treatment $t$, with $t$ being either 0 or 1. Thus, the Heterogeneous Treatment Effect (HTE) is expressed as $\tau(\mathbf{X}) = \mathbb{E}(Y(1) - Y(0) \mid \mathbf{X} ) $, which captures the expected treatment effect conditional on the covariates $\mathbf{X}$. We set $S = 0$ and $S = 1$ to denote OSs and RCTs, respectively. We aim to determine an estimator for $\tau(\mathbf{X})$, provided that $\mathbf{X}$ is within the range of values it can take in the RCTs. To maximize the efficiency benefits derived from the OSs, we proceed under the premise that the range of $\mathbf{X}$ values in the RCTs is either a subset of or intersects with the range in the OSs (\textit{overlap} assumption). This is because $\tau(\mathbf{X})$ cannot be identified for values of $\mathbf{X}$ that fall outside the scope of the RCTs. For brevity, within the subsequent text, "OS data" refers to observational data, while "RCT data" denotes randomized data.

Before introducing CIO, we present two common assumptions in causal inference \cite{johanssonLearningWeightedRepresentations2018,rosenbaumCentralRolePropensity1983,zhangLearningOverlappingRepresentations2020} and data fusion literature \cite{colnetCausalInferenceMethods2023,degtiarReviewGeneralizabilityTransportability2023,yangIntegrativeLearnerHeterogeneous2022}:

\begin{assumption}[Consistency, ignorability and overlap]\label{ass:basic}
For any individual $i$, assigned to treatment $t_i$, we observe $Y_i = Y (t_i)$. Further, $\{Y (t)\}_{t \in T}$ and the data generating process $p(\mathbf{X}, T, Y, S)$ satisfies strong ignorability: $T \perp \{Y(0), Y(1)\} \mid(\mathbf{X}, S=1)$ and overlap: $\forall x, 0<P(T \mid \mathbf{X})<1$.
\end{assumption}

\begin{assumption}[Transportability of the HTE]\label{ass:transable}
$\mathbb{E}(Y(1)-Y(0) \mid \mathbf{X}, S=s)=\tau(\mathbf{X}), s=0,1$.
\end{assumption}

The \textit{ignorability} assumption, one of the Assumption \ref{ass:basic}, often known as the no unmeasured confounders condition, which assumes that all variables influencing both the treatment $T$ and the outcome $Y$ are observed. It is inherently satisfied in an RCT owing to the nature of the random allocation of treatments. If Assumption \ref{ass:basic} holds, it is possible to determine the HTEs from RCT data. Conversely, in the case of OSs, the assumption of ignorable treatment assignment is not mandated, recognizing that such a requirement might be too stringent for practical applications.

As mentioned in \ref{method:pre}, we do not presume that the treatment assignment is ignorable in the context of OS data. Borrowed from \citet*{yangIntegrativeLearnerHeterogeneous2022}, under Assumptions \ref{ass:basic} and \ref{ass:transable}, the confounding bias function could be defined as the discrepancy between the conditional mean outcomes derived from OS data and the HTE:
\begin{align} \label{eq:eq1}
    c(\mathbf{X}) = \, & \mathbb{E}(Y \mid \mathbf{X}, T=1, S=0) - \\
    & \mathbb{E}(Y \mid \mathbf{X}, T=0, S=0) - \tau(\mathbf{X}) \nonumber
\end{align}
Adhering to the \textit{consistency} clause within Assumption \ref{ass:basic}, we have $\mathbb{E}(Y \mid \mathbf{X}, T=t, S=0) = \mathbb{E}(Y(t) \mid \mathbf{X}, T=t, S=0)$. Supposing the treatment assignment in the OS data is ignorable, namely $T \perp {Y(0), Y(1)} \mid(\mathbf{X}, S=0)$, it follows that $\mathbb{E}(Y(t) \mid \mathbf{X}, T=t, S=0) = \mathbb{E}(Y(t) \mid \mathbf{X}, S=0)$. Given Assumption \ref{ass:transable}, this would infer that the confounding bias function $c(\mathbf{X})$ is determined to be 0. That means, if the confounding bias do not exist within OS data, the confounding bias function $c(\mathbf{X})$ will be equal to zero.

We focus on a binary-treatment scenario, thus, for RCT data:
\begin{align}\label{eq:eq2}
    Y = \mathbb{E}(Y \mid \mathbf{X}, T=0, S=1) + T \tau(\mathbf{X}) + \epsilon_{r}.
\end{align}
The conditional expectation of residual $\epsilon_r$ equals to $\mathbb{E}(\epsilon_r \mid \mathbf{X}, T, S=1) = \mathbb{E}(Y \mid \mathbf{X}, T, S=1) - \mathbb{E}(Y \mid \mathbf{X}, T=0, S=1) - T \cdot \mathbb{E}(Y(1) - Y(0) \mid \mathbf{X}, S=1) $. According to the Assumption \ref{ass:basic} and \ref{ass:transable}, we obtain the following conclusion $\mathbb{E}(\epsilon_r \mid \mathbf{X}, T, S=1) = 0$. The proofs can be found in Appendix \ref{app:proofs}. Similarly, we formulate the outcome model for OS data based on Equation \ref{eq:eq1} and \ref{eq:eq2}:
 \begin{align}\label{eq:eq3}
     Y = \mathbb{E}(Y \mid \mathbf{X}, T=0, S=0) + T \cdot [\tau(\mathbf{X}) + c(\mathbf{X})] + \epsilon_{o}.
 \end{align}
Condition on Assumption \ref{ass:basic} and \ref{ass:transable}, we likewise derive the conclusion $\mathbb{E}(\epsilon_o \mid \mathbf{X}, T=0, S=0) = 0$. Its proofs are also demonstrated in Appendix \ref{app:proofs}. Therefore, when we combining OS data and RCT data for HTE estimation, we integrate Equation \ref{eq:eq2} and \ref{eq:eq3}:
\begin{align}\label{method:eq4}
    Y = \, & T [\tau(\mathbf{X}) + (1-S)c(\mathbf{X})] + \mathbb{E}(Y \mid \mathbf{X}, T=0, S) + \epsilon \\
    = \, & T [\tau(\mathbf{X}) + (1-S)c(\mathbf{X})] + \mu_0(\mathbf{X}) + \epsilon \nonumber,
\end{align}
where we set $\mu_0(\mathbf{X}) = \mathbb{E}(Y \mid \mathbf{X}, T=0, S)$. According to the conclusion $\mathbb{E}(\epsilon_r \mid \mathbf{X}, T, S=1) = 0$ and $\mathbb{E}(\epsilon_o \mid \mathbf{X}, T=0, S=0) = 0$, we can infer that the conditional expectation of $\epsilon$ equals to 0, \ie, $\mathbb{E}(\epsilon \mid \mathbf{X}, T, S) = 0 $.

\subsection{The Identification of Confounding Bias and HTEs} \label{method_pro}
Before introducing our method, we give the basic assumption which is deduced through Assumption \ref{ass:basic}:

\begin{assumption}[Basic assumption] \label{ass:our}
    Assuming $S\perp Y|X$ (meaning S and Y are independent given X), and combining the \textit{ignorability} and \textit{overlap} assumptions, we deduce that $T(1-S)\perp Y|X$ and $<P(T(1-S))<1$.
\end{assumption}

\vpara{Confounding bias estimation.}
Based on Equation \ref{method:eq4}, \citet*{yangIntegrativeLearnerHeterogeneous2022}  has created an integrative $R$-learner to estimate both the HTE and the confounding function. This learner harnesses randomized data for accurate identification while utilizing observational data to enhance its efficiency. However, such an approach necessitates comprehensive OS data, \ie, it requires data from both the treatment and control groups for learning the HTE estimator. As we previously stated, obtaining complete OS data in complex real-world scenarios is impractical. Therefore, there is a need for a robust HTE estimation method capable of data fusion, applicable to scenarios with complete OS data and, more crucially, adaptable to situations with incomplete OS data. For this purpose, we have developed a robust CIO to overcome the limitations of current data fusion techniques.

When integrating OS and RCT data, how to learn and eliminate confounding biases hidden in OS data is a critical procedure that we can not be circumvent. We decompose the Equation \ref{method:eq4} as follows,
\begin{align}\label{method:eq5}
    Y =  T(1-S)c(\mathbf{X}) + T \tau(\mathbf{X}) + \mu_0(\mathbf{X}) + \epsilon.
\end{align}
We define $D = T(1 - S)$ as an artificially generated treatment variable, where $D = 1$ is assigned when $T = 1$ and $S = 0$, and $D = 0$ otherwise. Following the creation of this proxy treatment mechanism, we form a pseudo-experimental group where $D = 1$ and a pseudo-control group where $D = 0$. Based on Assumption \ref{ass:our}, we the use the dummy data to learn the confounding bias function $c(\mathbf{X})$, of which the learning process is the same as effect estimation. Therefore, we denote the learned $c(\mathbf{X})$ as $\tau_c(\mathbf{X})$. It should be noted that the pseudo-control group comprises all the samples from the RCT dataset. 

\vpara{HTE estimation.}
After training the confounding bias function $\tau_c(\mathbf{X})$, we frozen the parameters of it. In the current stage, the Equation \ref{method:eq5} can be rearranged as:
\begin{align}\label{method:eq6}
    &\tilde{Y} = T \tau(\mathbf{X}) + \mu_0(\mathbf{X}) + \epsilon,\\
    &\tilde{Y} = Y - T(1 - S)\tau_c(\mathbf{X}) \nonumber,
\end{align}
where $T(1 - S)\tau_c(\mathbf{X}) $ is a constant for an individual. Following this equation, we then integrate the OS data with RCT data to train an effect estimator $\tau(\cdot)$. Utilizing this formula allows us to adjust and rectify the observed outcomes for the treatment group in the OS data. Thus, the OS data can be combined with RCT data to estimate HTEs without the inclusion of confounding biases. 

Diverging from existing approaches of data fusion for HTE assessment, the proposed CIO method does not require a propensity model to be learned beforehand, as seen in methods like FAST \cite{guFASTFusedAccurate2023} and the integrative $R$-learner \cite{yangIntegrativeLearnerHeterogeneous2022}, nor does it initially demand an effect estimator derived from OS data, which is a prerequisite for techniques such as RHC \cite{kallusRemovingHiddenConfounding2018} and CorNet \cite{hattCombiningObservationalRandomized2022}. The training process outlined above reveals the following insights:
\begin{itemize}[leftmargin=*]
    \item For estimating confounding biases, it is sufficient to use only the treated subset of OS data and RCT data.
    \item In estimating HTEs, even in the absence of access to the untreated group within the OS data, we have at our disposal, a composite treated group (encompassing both the treated individuals from the OS and the RCT) and a separate control group (consisting of the untreated units from the RCT). These groups can be employed to train an effect estimator.
\end{itemize}
These attributes endow the CIO method with resilience. In the above description of CIO, we assumed the control group of OS is missing for simplicity. \textit{However, CIO offers the versatility to adapt to situations where the treated cohort from OS data might be missing. This is achieved by inverting the treatment assignments for the treated and untreated samples, that is, assigning $T = 0$ for the initially treated subjects and $T = 1$ for the originally untreated subjects}.

\subsection{Training Loss}
We denote observational data as $\{\mathbf{x}_i^o, t_i^o, y_i^o\}_{i=1}^m$, thus its treated data and control data can be $ \{(\mathbf{x}_i^{ot}, y_i^{ot}) \mid t_i^o = 1 \}_{i=1}^{m_t}$ and $\{(\mathbf{x}_i^{oc}, y_i^{oc}) \mid t_i^o = 0 \}_{i=1}^{m_c}$, respectively. In the same way, we denote randomized data as $\{\mathbf{x}_i^r, t_i^r, y_i^r\}_{i=1}^n$, its treated data as $\{(\mathbf{x}_i^{rt}, y_i^{rt}) \mid t_i^r = 1 \}_{i=1}^{n_t}$ and its control data as $\{(\mathbf{x}_i^{rc}, y_i^{rc}) \mid t_i^r = 1 \}_{i=1}^{n_c}$. The $m, m_t, m_c, n, n_t, n_c$ represents the size of OS data, OS treated data, OS control data, RCT data, RCT treated data and RCT control data, respectively. 

\vpara{Stage 1: Confounding bias estimation. }
During this phase, various regression techniques can be utilized to model the data from individuals in OSs and RCTs, including ridge regression, random forest, neural networks, etc. Thus, we initialize $p_1(\cdot)$ and $p_0(\cdot)$ to correspondingly fit on pseudo-experimental data and pseudo-controls, \ie, the treated data of OSs and the all data of RCTs. According to Equation \ref{method:eq5}, the pertinent optimized objectives to this stage can be exemplified as follows:
\begin{align}
    \hat{p}_1(\cdot) & = \underset{p_1}{\operatorname{argmin}} \frac{1}{m_t} \sum_{i=1}^{m_t}[y_i^{ot} - p_1(\mathbf{x}_i^{ot})]^2, \\
    \hat{p}_0(\cdot) & = \underset{p_0}{\operatorname{argmin}} \left\{ \frac{1}{n_t} \sum_{i=1}^{n_t}[y_i^{rt} - p_0(\mathbf{x}_i^{rt})]^2 + \frac{1}{n_c} \sum_{i=1}^{n_c}[y_i^{rc} - p_0(\mathbf{x}_i^{rc})]^2 \right\}.
\end{align}
After that, we calculate the confounding bias function $\hat{\tau}_c(\mathbf{x}_i) = \hat{p}_1(\mathbf{x}_i) - \hat{p}_0(\mathbf{x}_i)$ for each unit from OS treatment group.

\vpara{Stage 2: HTE estimation. }
At this stage, we amalgamate the entire dataset from OSs with that from RCTs while also adjusting the observed outcomes for the treated group in OSs using the estimated confounding bias function $\tau_c(\cdot)$. In a similar vein, we set up the functions $f_1(\cdot)$ and $f_0(\cdot)$ to be trained on the entirety of the treatment and control data, drawn from the aggregate of OSs and RCTs, correspondingly. It should be noted that for effectively debiasing with $\hat{\tau}_c(\cdot)$, we adopt the parameters from $\hat{p}_1(\cdot)$ as the initial values for $f_1(\cdot)$. Additionally, it is imperative to conduct initial training for $f_0(\cdot)$ using control data, ensuring that the training epochs align with those of the stage 1. According to Equation \ref{method:eq6}, the loss function is presented as follows:
\begin{align}
    \hat{f}_1(\cdot) & = \underset{f_1}{\operatorname{argmin}} \left\{ \frac{1}{m_t} \sum_{i=1}^{m_t}\{\tilde{y}i^{ot} - f_1(\mathbf{x}_i^{ot})\}^2 + \frac{1}{n_t} \sum_{i=1}^{n_t}(y_i^{rt} - f_1(\mathbf{x}_i^{rt}))^2 \right\},  \\
    \hat{f}_0(\cdot) & = \underset{f_0}{\operatorname{argmin}} \left\{ \frac{1}{m_c} \sum_{i=1}^{m_c}[y_i^{oc} - f_0(\mathbf{x}_i^{oc})]^2 + \frac{1}{n_c} \sum_{i=1}^{n_c}[y_i^{rc} - f_0(\mathbf{x}_i^{rc})]^2 \right\},
\end{align}
where $\tilde{y}_i^{ot} = y_i^{ot} - \hat{\tau}_c(\mathbf{x}_i^{ot})$. Finally, we obtain HTE estimator $\hat{\tau}(\mathbf{x_i}) = \hat{f}_1(\mathbf{x_i}) - \hat{f}_0(\mathbf{x_i})$ for each unit.

\subsection{Effectiveness in Theoretical}
In this paper, following the approach of intergrative $R$-learner \citet{yangIntegrativeLearnerHeterogeneous2022}, we introduce the confounding function $c$ to describe the confounding bias in Observational Studies (OS), as shown by the Equation \ref{method:eq5} $Y = T(1-S)c(X)+T\tau(X)+\mu_0(X)+\epsilon$. When treating $T(1-S)$ as a dummy treatment variable, we are able to estimate $c(X)$, which constitutes the first stage in our paper. Assuming $S\perp Y|X$ (meaning S and Y are independent given X), and combining the \textit{ignorability} and \textit{overlap} assumptions, we deduce that $T(1-S)\perp Y|X$ and $<P(T(1-S))<1$. This implies that, under the Potential Outcome Framework (POF), $c$ is identifiable in theoretical. Since the $c$ is identified under POF, we denote the identified $c$ as $\tau_c$. After acquiring the confounding bias function $\tau_c$, we calibrate the outcome of OS using it to reduce the confounding bias. Finally, in the second stage, HTE estimation is generally based on the POF framework and is also identifiable. In summary, we decompose the process of HTE estimation combining OS and RCT data into two stages under POF, which are guaranteed by the effectiveness of POF.

\section{Experiment }
In this section, we perform experiments on a synthetical dataset and two real-world datasets to demonstrate the performance of CIO for HTE estimation. The outcome of these data are all simulated by a certain strategy. We present the results of various experiments designed to address the subsequent three research questions:
\begin{itemize}[leftmargin=*]
     \item \textbf{RQ1:} Wether our proposed approach CIO is effective to combine observational data and randomized data for HTE estimation under the two situation: observational data is complete or incomplete?
    \item \textbf{RQ2:} Should the strength of confounding bias presented in OS data intensify, would the proposed approach maintain its superior performance relative to current data fusion techniques?
    \item \textbf{RQ3:} The impact of inverted treatment assignment on HTE estimation.
\end{itemize}
The following text will include three subsections: Experimental Setup, Datasets, and Experimental Analysis. Within the Experimental Analysis subsection, we conduct a mass of experiments and the corresponding analysis to answer the above research questions.

\subsection{Experimental Setup}
\vpara{Baselines and architectures.}
To assess the performance of CIO, we choose RHC \cite{kallusRemovingHiddenConfounding2018}, integrative $R$-learner \cite{yangIntegrativeLearnerHeterogeneous2022} and CorNet \cite{hattCombiningObservationalRandomized2022} as baselines for comparison. Following RHC \cite{kallusRemovingHiddenConfounding2018}, we select Ridge and RF as our base model for comparison. In addition, given that methods based on representation learning have demonstrated notable efficacy in estimating heterogeneous treatment effects (HTEs), we follow the approach of CFR \cite{shalitEstimatingIndividualTreatment2017} by integrating the Treatment-Agnostic Representation Network (TARNet) into our suite of baseline models for experimental purposes. For the sake of brevity in descriptions, we refer to the integrative $R$-learner as the IntR. RHC, IntR, CorNet and CIO all follow a dual-phase training approach that leverages OS and RCT data for estimating HTE and are implemented using ridge regression (Ridge), random forest (RF) and TARNet as underlying models. Since CorNet is a method implemented with DNN network including representation layer, we only compare with it when we implement other baselines and CIO with TARNet. Furthermore, to determine the benefits of combining OS and RCT data, we train baseline estimators solely on each data type, referred to as SF$_{OS}$ for OS data and SF$_{RCT}$ for RCT data. We also conduct an experiment where OS and RCT data are simply fused for training to serve as a reference for the effectiveness of modeling confounding bias, indicated as SI.

\vpara{Incomplete OS data.}
Most importantly, the original purpose of CIO is to combine the OS and RCT data for HTE  estimation in the situation where the OS data is not fully available. Therefore, in order to create the incomplete condition, we deliberately omit either the untreated or treated group from the OS datasets of the Simulation and STAR experiments during the training phase. This modified approach is referenced as CIO$_{IO}$.



\vpara{Metrics.}
When developing models to predict the Individual Treatment Effect (ITE), the main objective is to minimize the Precision in the Estimation of Heterogeneous Effect (PEHE) as outlined in reference \cite{hillBayesianNonparametricModeling2011,schwabPerfectMatchSimple2019}. In a binary treatment scenario, PEHE quantifies the accuracy with which a model can predict the differential impact of two treatments, $t_0$ and $t_1$, for a given set of samples $X$. To calculate PEHE, we determine the mean squared error across $N$ samples by comparing the actual difference in outcomes, $y_1(n) - y_0(n)$, which are obtained from the simulation strategy, with the predicted difference, $\hat{y}_1(n) - \hat{y}_0(n)$, where $n$ denotes the sample index:
\begin{align}
\epsilon_{\mathrm{PEHE}}=\frac{1}{N} \sum_{n=0}^N\left(\left[y_1(n)-y_0(n)\right]-\left[\hat{y}_1(n)-\hat{y}_0(n)\right]\right)^2
\end{align}

\begin{table}[t]
    \centering
    \caption{Comparison of methods for combining OS and RCT data on Simulation, STAR and NSW data. We report the mean value $\pm$ the standard deviation of $\sqrt{\epsilon_{PEHE}}$ on test data over 10 repeated runs for three data with proportion $p_{r} = 0.2$ of RCT data, respectively. Besides, we remove the control group of OS in Simulation and STAR datasets and present their results in $\textbf{CIO}_{IO}$. We run trials for 10 times and the best performance is marked in \textbf{bold}. }
    \resizebox{\linewidth}{!}{
    \begin{tabular}{c|c|ccc}
    \toprule
    \multirow{2}{*}{Architecture}  & \multirow{2}{*}{Method} & \multicolumn{3}{c}{$p_r=0.2$} \\
    \cmidrule{3-5}
    & & Simulation & STAR & NSW \\
    \midrule
    \multirow{7}{*}{Ridge} 
    & SF$_{OS}$ & 21.97 $\pm$ 1.06 & 25.05 $\pm$ 2.92 & -  \\
    & SF$_{RCT}$ & 9.74 $\pm$ 2.39 & 4.06 $\pm$ 0.52 & 2.49 $\pm$ 1.06 \\
    & SI & 21.43 $\pm$ 1.06 & 12.82 $\pm$ 3.48 & 15.85 $\pm$ 0.32 \\
    & RHC & 14.62 $\pm$ 6.18 & 9.7 $\pm$ 2.68 & - \\
    & IntR & 7.68 $\pm$ 0.34 & 8.52 $\pm$ 0.71 & - \\
    \cmidrule{2-5}
    & \textbf{CIO}(Ours) & \textbf{5.75} $\pm$ 0.34  & 2.36 $\pm$ 0.48 & - \\
    & \textbf{CIO$_{IO}$}(Ours) & 8.96 $\pm$ 0.63 & \textbf{2.14} $\pm$ 0.65 & \textbf{1.48} $\pm$ 0.38 \\
    
    \midrule
    \multirow{7}{*}{RF} 
    & SF$_{OS}$ & 18.35 $\pm$ 0.67 & 48.17 $\pm$ 0.47 & - \\
    & SF$_{RCT}$ & 10.41 $\pm$ 2.87 & 6.96 $\pm$ 1.30 & 2.30 $\pm$ 0.70 \\
    & SI & 17.83 $\pm$ 0.73 & 7.83 $\pm$ 0.87 & 3.84 $\pm$ 1.08 \\
    & RHC & 11.76 $\pm$ 1.28 & 18.59 $\pm$ 0.50 & - \\
    & IntR & 8.84 $\pm$ 1.93 & 9.51 $\pm$ 0.22 & - \\
    \cmidrule{2-5}
    & \textbf{CIO}(Ours) & \textbf{6.65} $\pm$ 0.26 & 5.61 $\pm$ 0.93 & - \\
    & \textbf{CIO$_{IO}$}(Ours) & 10.18 $\pm$ 2.94 & \textbf{5.35} $\pm$ 0.96 & \textbf{2.29} $\pm$ 0.70 \\
    
    \midrule
    \multirow{6}{*}{TARNet} 
    & SF$_{OS}$ &  23.64$\pm$1.30  & 30.40$\pm$11.62 & - \\
    & SF$_{RCT}$ &  10.84$\pm$6.41  & 5.15 $\pm$ 3.38 & 5.14 $\pm$ 0.46 \\
    & SI &  22.83$\pm$1.08 & 23.81 $\pm$ 7.96 & 22.47 $\pm$ 0.60 \\
    & RHC &  7.89$\pm$3.40  & 7.06$\pm$ 2.65 & - \\
    & CorNet &  12.24$\pm$2.25  & 7.00$\pm$ 1.01 & - \\
    & IntR &  6.97$\pm$0.72  & 4.93$\pm$ 1.81 & - \\
    \cmidrule{2-5}
    & \textbf{CIO}(Ours) &  \textbf{6.61}$\pm$0.35  & \textbf{3.54} $\pm$ 1.38 & - \\
    & \textbf{CIO$_{IO}$}(Ours) &  6.92$\pm$0.41  & 4.17 $\pm$ 1.39 & \textbf{5.11} $\pm $ 0.39 \\
    \bottomrule
    \end{tabular}}
    \label{tab1}
\end{table}

\subsection{Datasets} \label{subsec:sim_study}
In line with prior approaches \cite{guFASTFusedAccurate2023,hattCombiningObservationalRandomized2022,yangImprovedInferenceHeterogeneous2022,yangIntegrativeLearnerHeterogeneous2022}, we create a simulated dataset and choose two real-world datasets, STAR and NSW, for our experimental evaluations. Due to the absence of ground truth for the effects in STAR and NSW, we simulate the outcomes for these datasets rather than relying on their actual outcomes. In this subsection, we will introduce more details of three experimental datasets about their data construction and outcome simulation.

\subsubsection{Simulation Dataset} A synthetic dataset, of which the covariates and outcomes are all simulated:
\begin{itemize}[leftmargin=*]
    \item \vpara{Data construction and outcome simulation.} 
    We generate independent covariates of dimension $p$, denoted as $X_i$, from a standard normal distribution $X_i \sim \mathcal{N}(0, 1) $ for $ i = 1, 2, \ldots, p $. In this experiment, we set $p=5$. Following this procedure, we produce 200 samples for RCT data, 3000 for OS data, and 1000 for test data. The potential outcomes for each sample are then simulated using the equation $Y(t) = t\tau(\mathbf{X}) + 1 + 2\sum_{i=1}^p X_i^3 + \sum_{i=1}^p X_i + 5(1 - s)U + \epsilon(t)$ , where $\tau(\mathbf{X}) = 1 + \sum_{i=1}^p X_i + \sum_{i=1}^p X_i^2$, $t \in \{0, 1\}$ indicates the treatment status, $s$ equalling 0 refers to OS data and 1 to RCT data, and $\epsilon(t)$ is drawn from $\mathcal{N}(0, 1)$. Treatment assignment for RCT and OS data is governed by $T|(\mathbf{X}, S=1) \sim Bernoulli(0.5)$ and $T|(\mathbf{X}, S=0) \sim Bernoulli(1 / (1 + \exp(-\sum_{i=1}^p X_i)))$, respectively. Echoing the simulation approach in \cite{yangImprovedInferenceHeterogeneous2022}, the unobserved variable $U$ is sampled from $\mathcal{N}(\mathbf{X}^T \mathbf{v} \beta (2T-1), 1)$, with $\mathbf{v}$ being a unit vector $(1, \ldots, 1)^T$ and $\beta$ being a coefficient that modulates the magnitude of confounding bias in OS data.
\end{itemize}

\subsubsection{STAR} A semi-synthetic dataset. Beyond synthetic dataset experiments, we also evaluate the performance of CIO using a real-world dataset in this subsection:
\begin{itemize}[leftmargin=*]
    \item \vpara{Tennessee Student/Teacher Achievement Ratio (STAR).}
    The STAR Experiment \cite{kruegerExperimentalEstimatesEducation1999} was a randomized controlled trial conducted in the late 1980s. Its objective is to measure the impact of class size on students' academic performance. We follow RHC \cite{kallusRemovingHiddenConfounding2018} and FAST \cite{guFASTFusedAccurate2023} to split STAR for getting observational and randomized data. Our attention is centered on two experimental classroom size conditions: small classes consisting of 13-17 students and regular classes with 22-25 students. Considering that a significant number of students commenced the study beginning in the first grade, we designate the type of class they were placed in at that time as their initial treatment. For each student, we consider a set of variables: gender, race, birth month, birthday, birth year, free lunch given or not, teacher id. We exclude any students who have missing data for any of these specified covariates. Furthermore, we also eliminate students whose combined scores for standardized tests in listening, reading, and math are missing. In total, we recorded 4139 students: 1774 assigned to treatment (small class, T = 1), and 2365 to control (regular size class, T = 0).
    \item \vpara{Data construction.}
    Through a specific simulation strategy, we get the outcome for each student. Then, we follow the settings of RHC \cite{kallusRemovingHiddenConfounding2018} and FAST \cite{guFASTFusedAccurate2023} to construct OS data, RCT data and test data: To introduce a confounding bias, we divide the study population based on a variable: students living in rural or inner-city areas (denoted by U = 1, totaling 2811 students) versus those in urban or suburban areas (denoted by U = 0, totaling 1407 students). We then create the RCT data by randomly selecting a proportion, which equals 0.5, of the students with U = 1. The OS data is compiled in the following manner: For students with U = 1, we include those who are not part of the trial data and have a treatment status of control (D = 0), along with the treated students (D = 1) whose simulated outcomes were in the bottom 50\% among their peers with D = 1 and U = 1. For students with U = 0, we incorporate all of the control students (D = 0) and the treated students (D = 1) who also fell into the bottom 50\% of simulated outcomes for the group with D = 1 and U = 0. Finally, the test data is composed of a reserved subset of the entire sample, excluding those individuals that are included in the RCT dataset.
    \item \vpara{Outcome simulation.}
    Following RHC \cite{kallusRemovingHiddenConfounding2018} and FAST \cite{guFASTFusedAccurate2023}, the actual covariates $\mathbf{X} = (X_1, X_2, \cdots, X_p)^T$, where $p = 7$. While the STAR dataset contains actual outcome data, it lacks a ground truth for the treatment effect. Consequently, we simulate both the outcome and the treatment effect function specifically for the STAR dataset. Concretely, we set $Y(t) = t\tau(\mathbf{X}) + 2\sum_{i=1}^p X_i + \mathbf{X}^T \mathbf{X} + \epsilon(t) $ as potential outcomes, where $\tau(\mathbf{X}) = \sum_{i=1}^p X_i +  \sqrt{|\sum_{i=1}^p X_i|} $ and $\epsilon(t) \sim \mathcal{N}(0, 1) $.
\end{itemize}

\subsubsection{NSW} A semi-synthetic dataset with incomplete OS dataset. The datasets illustrated in the previous subsections include complete OS data. While we have shown in Table \ref{tab1} that CIO maintains its robustness when faced with incomplete OS data, achieved by omitting the control group of OSs, this does not fully convey the method's practical utility in an intuitive manner. In light of this, we evaluate on inherently incomplete dataset --- NSW, where the treated data of OS data is absent:
\begin{itemize}[leftmargin=*]
    \item \vpara{National Supported Work (NSW) Demonstration. }
    The National Supported Work (NSW) Demonstration \cite{lalondeEvaluatingEconometricEvaluations1986a} was a randomized experiment investigating the effect of job training on income and employment status. Following \cite{a.smithDoesMatchingOvercome2005}, we combine randomized samples (297 treated, 425 control) with the 2490 PSID observational controls in our experiments.
    \item \vpara{Data construction.}
    The incomplete nature of the NSW's OS dataset makes it an apt choice for testing the resilience of the CIO method. Thus, we randomly draw 100 samples, encompassing both the treated group and the control group, from the pool of 722 randomized samples to serve as RCT data. To inject additional bias into the OS data, we incorporate those samples with simulated outcomes ranking in the upper 50\% from the 2490 PSID observational controls. The rest of the randomized samples and observational controls are allocated for the testing phase.
    \item \vpara{Outcome simulation.}
    We use the actual covariates $\mathbf{X} = (X_1, X_2, \cdots \\, X_p)^T$ of NSW: age, level of education, ethnicity (split into two covariates), marital status, and educational degree, where $p=6$. We generate potential outcomes for each person by $Y(t) = t \tau(\mathbf{X}) + 2 \sum_{i=1}^p \exp(X_i) + \epsilon(t)$, where $\tau(\mathbf{X}) = \mathbf{X}^T \mathbf{X} $, $t = \{0, 1\} $ and $\epsilon(t) \sim \mathcal{U}(-1, 1)$. It should be noted that since the treat data of OS data is inexistent, we set the treatment value of RCT treated data to 0, the control data of OS and RCT to 1.
\end{itemize}

\subsection{Experiment Analysis}
\begin{figure}[t]
\centering
\subfigure[Simulation dataset]{
		\includegraphics[width=\linewidth]{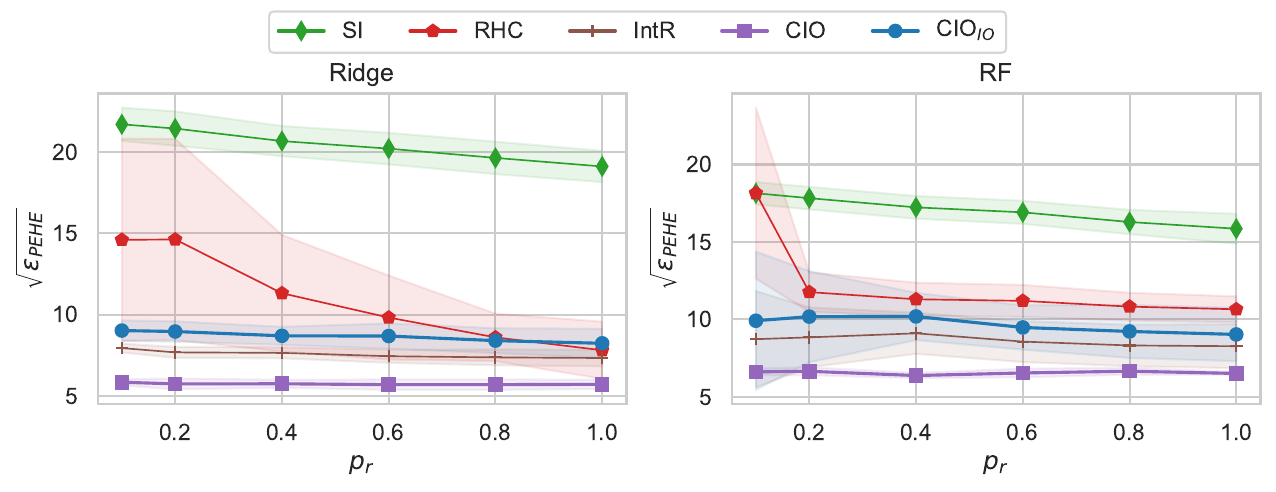}
            \label{fig1_a}
  }\\
\subfigure[STAR dataset]{
		\includegraphics[width=\linewidth]{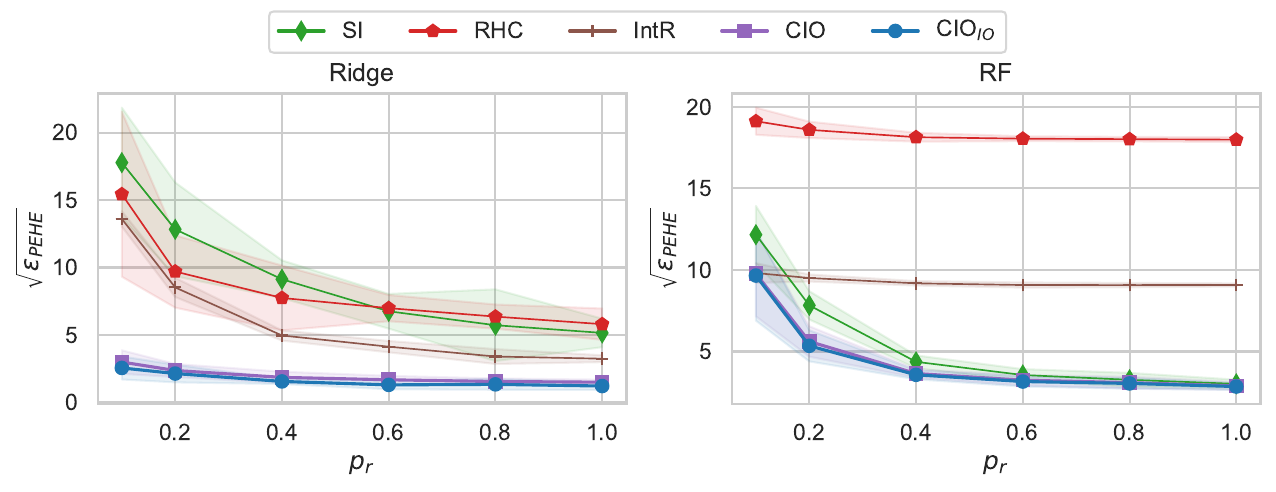}
            \label{fig1_b}
}\\
\subfigure[NSW dataset]{
		\includegraphics[width=\linewidth]{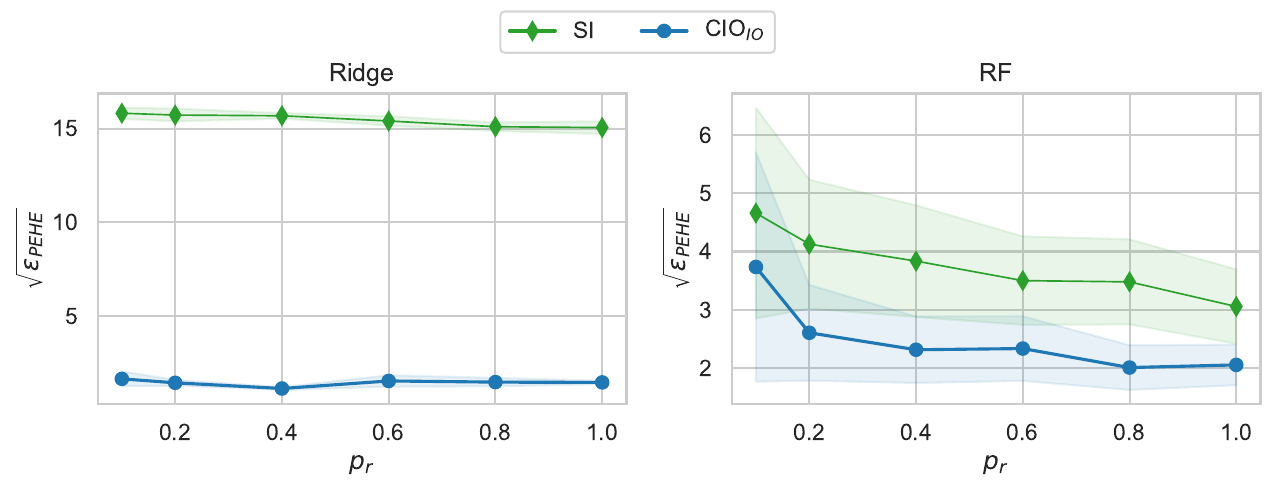}
            \label{fig1_c}
  }
  \vspace{-0.2cm}
\caption{Comparison among data-fusion baselines under Ridge and RF with an increasing ratio of RCT data for training. We plot the results upon Simulation dataset, STAR dataset and NSW dataset on Figure 1(a), 1(b) and 1(c) respectively.}
\label{fig1}
\vspace{-0.3cm}
\end{figure}

\vpara{Preliminary Trials.}
Initially, we execute CIO alongside other benchmark methods using Ridge, RF, and TARNet to calculate HTEs, an essential functionality of these methodologies. Our investigation aims to determine if integrating even a tiny quantity of RCT data with OS data yields any advantages for HTE estimation. To this end, we randomly select a subset of the RCT data, maintaining a proportion $p_r=0.2$, and combine it with the entirety of the OS data for the training process. We promise that the selected RCT data includes treated and control instances. This strategy is designed to reflect the common scenario encountered in the real world, where RCT data is often considerably less abundant than OS data. The outcomes of this analysis are compiled and presented in Table \ref{tab1}. It can be observed that relying solely on OS data during the training phase introduces significant bias, resulting in poor performance outcomes. Likewise, even though SI integrates OS data with RCTs, it only shows a marginal improvement over SF$_{OS}$. This is because SI merges OS data and RCT data directly without implementing any procedures to mitigate bias. In pursuit of this objective, RHC and IntR are designed to correct for biases in OS data by leveraging RCT data in the estimation of HTEs. The results presented in Table \ref{tab1} indicate that CIO significantly outperforms IntR when complete OS data is used, highlighting CIO's effectiveness in mitigating confounding bias from OS data. For examples, we have conducted a significance test between IntR and CIO On Simulation dataset, the p-value is 4.27e-10 for Ridge, 3.43e-3 for RF; On STAR dataset, the p-value is 1.79e-6 for Ridge, 7.83e-3 for RF.  Another important aspect of our analysis involves the removal of control group data from the OS set and merging the remaining OS data with the full RCT dataset to assess CIO's resilience. Even when faced with incomplete OS data, CIO is capable of leveraging the available OS data for training purposes. Despite a noticeable decline in the performance of CIO$_{IO}$ as compared to the complete CIO, it still surpasses SF$_{RCT}$ in terms of effectiveness. This underscores CIO's ability to preserve a robust performance in HTE estimation by making use of the available data. Moreover, as previously demonstrated, both RHC and IntR are incapable of integrating RCT with OS data in the presence of partial missingness in the OS dataset. As a result, we are limited to only assessing the outcomes of CIO$_{IO}$, SI, and SF$_{RCT}$ solely on the NSW dataset. Employing Ridge as the underlying model, our proposed approach attains superior performance. The significance test yields a p-value of 0.01, falling below the threshold of 0.05. When RF is employed as the underlying architecture, the performance of CIO$_{IO}$ is on par with that of SF$_{RCT}$. 

\vpara{Sensitivity of RCT data volume.} 
In the previously mentioned experiments, CIO demonstrated its efficacy and resilience when provided with a small amount of RCT data for training. Yet, it remains unclear if it can sustain advanced performance when supplied with varying volume of RCT data for training. To investigate this, we modify the proportion of RCT data used in training---$p_r$, ranging from 0.1 to 1.0. The results of all methods implemented with Ridge regression, RF are shown in Figure \ref{fig1}. We summarize three aspects from the figure: 
\begin{itemize}[leftmargin=*]
    \item Intuitively, the $\sqrt{\epsilon_{PEHE}}$ results of all methods will decrease with the size of RCT data swells, given that RCT data are devoid of unobserved confounders. Our approach consistently surpasses alternative data-fusion techniques when applied with Ridge and RF models, thereby affirming the efficacy of CIO in estimating HTEs.
    \item More importantly, to assess the robustness of CIO in scenarios where the OS data are incomplete, we conduct experiments without the controls from OS data. Results indicate that CIO$_{IO}$ maintains superior performance, even in the face of partial absence of OS data.
    \item Under the Ridge model, RHC and IntR demonstrate enhanced performance compared with SI, whereas this performance edge is not observed with RF. In contrast, CIO consistently outperforms SI regardless of whether Ridge or RF is employed, highlighting CIO's stable debiasing capability.An observation of the figure reveals that CIO consistently secures the top performance tier, regardless of the amount of RCT data utilized.
\end{itemize} 
Furthermore, it is notable that the standard deviation values in Figure 1 significantly vary across different data ratios within the same dataset. At lower ratios, only a minimal amount of RCT data is provided for training, which can result in substantial variance in the data samples used for training in different experimental runs, consequently leading to higher variance in performance. As the volume of RCT data incorporated into training increases, the fluctuation in the training data set decreases, leading to performance stability throughout multiple experimental trials.

\vpara{Impact of the strength of confounding bias.}
\begin{figure}[t]
    \centering
    \includegraphics[width=0.78\linewidth]{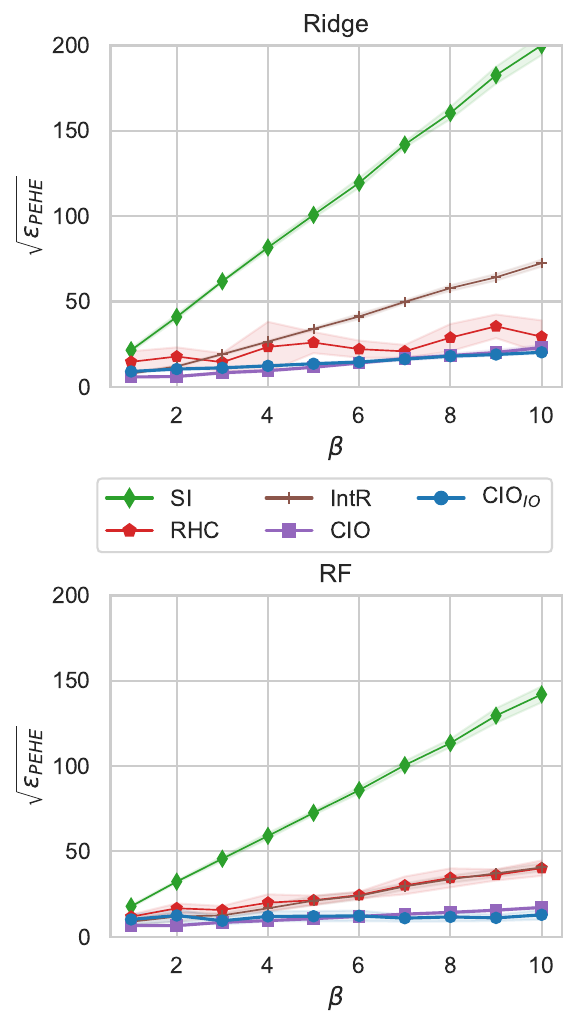}
    \vspace{-0.2cm}
    \caption{For all data-fusion techniques using Ridge and RF, we observe $\sqrt{\epsilon_{PEHE}}$ across a range of $\beta$ values that modulate the intensity of the confounding bias in the training OS data. }
    \vspace{-0.5cm}
    \label{fig:bias}
\end{figure}
Our approach involves identifying the confounding bias present in OS data and using this as a residual adjustment for the observed outcomes. Recognizing the confounding bias in OS data is essential when integrating OS and RCT data. To this effect, we manipulate the intensity of the confounding bias in OS data by varying the $\beta$ value. The results of this manipulation are illustrated in Figure \ref{fig:bias}, where the OS data will suffer from enhancing confounding bias with the value of $\beta$ rising. It is evident that the efficacy of SI declines sharply as $\beta$ increases, whereas the performance of other data-fusion methods deteriorates at a more gradual rate, which demonstrates the effectiveness of identifying the confounding bias. Particularly, CIO can maintain a low $\sqrt{\epsilon_{PEHE}}$ despite high values of $\beta$, regardless of whether the OS data is complete. It should be highlighted that as $\beta$ reaches a specific threshold, the performance of CIO$_{IO}$ surpasses that of CIO. This phenomenon may be attributed to the trade-off between the quantity of OS control data and the strength of associated confounding bias. We can exploit the characteristic when OS data lacks control or treatment data and suffers from substantial confounding bias.

\begin{figure}[t]
    \centering
    \subfigure[Simulation dataset]{
		\includegraphics[width=0.8\linewidth]{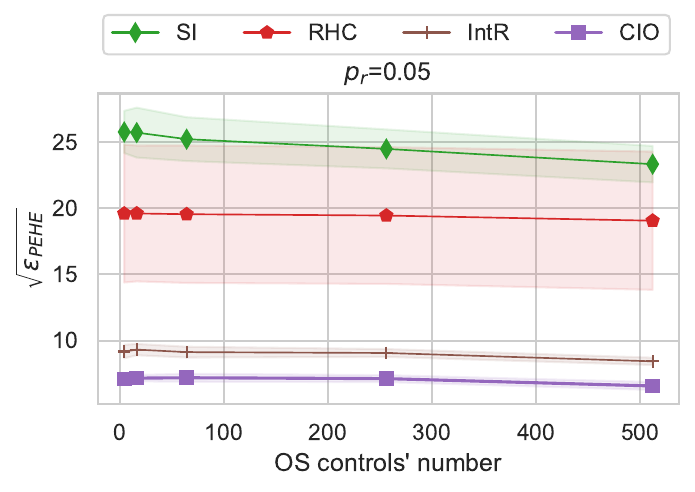}
            \label{fig:os_control_a}
    }
    \subfigure[STAR dataset]{
		\includegraphics[width=0.8\linewidth]{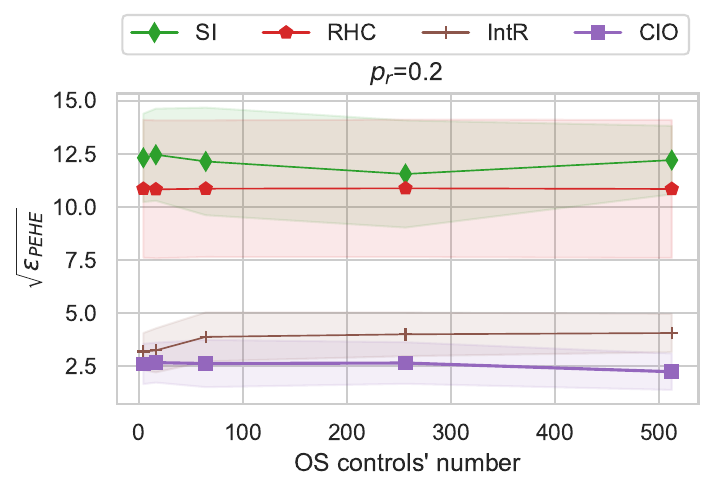}
            \label{fig:os_control_b}
    }
    \caption{ We change the quantity of control data from the OS used in the training stage, under which we evaluate the efficacy of various data-fusion techniques implemented with \textit{Ridge} regression. The OS controls' number varies from a range of \{1, 4, 16, 64, 256, 512\}. Results pertaining to the Simulation dataset are illustrated in Figure 3(a) and for the STAR dataset in Figure 3(b).}
    \vspace{-0.3cm}
    \label{fig:os_control}
\end{figure}

\vpara{Impact of the size of OS control data.}
In scenarios where the OS control group data is nonexistent, CIO remains unique ability of integrating the available OS data with RCT data for estimating HTE. Consequently, it becomes intriguing to examine the fluctuating performance of both baseline methods and CIO when trained with varying volumes of OS control data. We randomly select RCT data with $p_r=0.05$(ensuring the inclusion of treatment and control samples) for training. The results are displayed in Figure \ref{fig:os_control_a}. CIO consistently surpasses alternative methods in performance across all sizes of OS control data, demonstrating the applicability of CIO for handling data with varying and intricate compositions.
Following the trials conducted for the Simulation dataset, we change the number of OS control data from a same range. To generate a distinct composition of RCT data for the training set, we set $pr=0.2$, as depicted in \ref{fig:os_control_b}. According to the line chart, CIO maintains a steady enhancement in performance over other baseline methods, irrespective of the OS control data size. This illustrates CIO's exceptional robustness.

\vpara{Inverse of Treatment Assignment.}
In Section \ref{method}, for the convenience of describing CIO, we assume that the control group of OSs is unavailable. Nevertheless, in practical situations, it's often the treatment group's data that might be missing. As previously detailed, in the absence of treatment data, we can reverse the original treatment assignments to adapt our methodology. This reversal process enables the flexible application of CIO for merging both OS and RCT data to estimate HTEs. While this approach provides flexibility for CIO, its efficacy post-inversion remains to be seen. To address this, we perform a series of experiments to assess CIO's robustness. Initially, we eliminate the treatment group of OSs from the Simulation and STAR datasets for our analysis. Subsequently, we extend our experiments by removing and inverting the treatment assignments within the OS data and RCT data of these datasets to further examine the performance of the proposed method. We report the experimental results in Table \ref{tab2}, where the performance of CIO is almost invariable between 'original' and 'inverse'. Such consistency highlights that CIO retains its effectiveness even in cases where the treatment group data from OSs is absent. 

\begin{table}[t]
    \centering
    \caption{For both the Simulation and STAR datasets, we exclude the control group from the OS data for experimenting. Conversely, the treatment group of OS data is eliminated from the Simulation dataset and from STAR, with their treatment assignments being inverted. The 'original' means the original treatment assignment, while the 'inverse' represents the inverted treatment assignment. }
    \resizebox{\linewidth}{!}{
    \begin{tabular}{c|cc|cc}
    \toprule
    \multirow{2}{*}{Base Model}  & \multicolumn{2}{c}{Simulation} & \multicolumn{2}{c}{STAR} \\
    & original & inverse & original & inverse \\
    \midrule
    Ridge & 8.96$\pm$0.63 & 8.03$\pm$0.51 & 2.14$\pm$0.65 & 2.65$\pm$0.99 \\
    RF & 10.18$\pm$2.94 & 9.79$\pm$2.52 & 5.35$\pm$0.96 & 5.24$\pm$0.61 \\
    TARNet & 6.92$\pm$0.41 & 7.12$\pm$0.29 & 4.17$\pm$1.39 & 4.24$\pm$2.19 \\
    \bottomrule
    \end{tabular}}
    \vspace{-0.5cm}
    \label{tab2}
    
\end{table}

\section{Conclusion}
This paper posits that existing data-fusion techniques are deficient in robustness, rendering them incapable of merging OS data with RCT data for HTE estimation in instances where the OS training data is incomplete. In response to this issue, we present CIO, a resilient method designed to harness the advantages of both OS and RCT data. CIO circumvents the limitations of current data fusion methods by effectively estimating HTEs without requiring fully populated OS datasets. To achieve this, we form pseudo-experimental and pseudo-control groups from another perspective to train an estimator for effect measurement which is intended to serve as a confounding bias function, an innovative tactic in assessing confounding bias. To confirm the robustness and effectiveness of our method, we perform numerous tests that explore various aspects: we examine the influence of RCT data volume, analyze the effect of confounding bias intensity, and investigate how the quantity of OS control data affects outcomes. These trials are conducted on a synthetic dataset and two semi-synthetic datasets which use real-world covariates and outcomes generated via specific strategies. Across all experiments, CIO's performance consistently surpasses that of the baseline methods it is compared with, irrespective of the dataset and architecture used. The consistent outperformance of CIO when benchmarked against other data-fusion methods affirms the effectiveness and robustness of our confounding bias estimator. This tool, which calibrates the observed outcomes of OS data, proves to be powerful in merging RCT and OS data for estimating HTEs.

\appendix
\section{Proofs} \label{app:proofs}

\vpara{Proof1.} 
$\mathbb{E}( \epsilon_{r} \mid \mathbf{X}, T=0, S=1) = \mathbb{E}(Y \mid \mathbf{X}, T=0, S=1) -\\
\mathbb{E}(Y \mid \mathbf{X}, T=0, S=1) = 0.$

\vpara{Proof2.}
$
\mathbb{E}(\epsilon_r \mid \mathbf{X}, T=1, S=1) = \mathbb{E}(Y \mid \mathbf{X}, T=1, S=1)
- \mathbb{E}(Y \mid \mathbf{X}, T=0, S=1) - \mathbb{E}(Y(1) - Y(0) \mid \mathbf{X}, S=1) 
= \mathbb{E}(Y(1) \mid \mathbf{X}, T=1, S=1) - \mathbb{E}(Y(0) \mid \mathbf{X}, T=0, S=1)  - 
\mathbb{E}(Y(1) - Y(0) \mid \mathbf{X}, S=1) = \mathbb{E}(Y(1) \mid \mathbf{X}, S=1) 
\mathbb{E}(Y(0) \mid \mathbf{X}, S=1) - \mathbb{E}(Y(1) - Y(0) \mid \mathbf{X}, S=1) = 0. 
$

\vpara{Proof3.}
$
\mathbb{E}(\epsilon_o \mid \mathbf{X}, T=0, S=0) = \mathbb{E}(Y \mid \mathbf{X}, T=0, S=0) -\\
\mathbb{E}(Y \mid \mathbf{X}, T=0, S=0) = 0.
$

\vpara{Proof4.}
$
\mathbb{E}(\epsilon_o \mid \mathbf{X}, T=1, S=0) = \mathbb{E}(Y \mid \mathbf{X}, T=1, S=0) -
\mathbb{E}(Y \mid \mathbf{X}, T=0, S=0) - [\mathbb{E}(Y \mid \mathbf{X}, T=1, S=0) - 
\mathbb{E}(Y \mid \mathbf{X}, T=0, S=0)] = 0.
$ 
\bibliographystyle{ACM-Reference-Format}
\bibliography{ref} 

\end{document}